\documentclass[12pt,a4paper]{iopart}
\usepackage[breaklinks]{hyperref}
\usepackage[latin1]{inputenc}
\usepackage{graphicx}
\usepackage{color}
\usepackage{amssymb}
\usepackage{bm,bbm}
\usepackage[OT2,OT1]{fontenc}
\newcommand\cyr{%
\renewcommand\rmdefault{wncyr}%
\renewcommand\sfdefault{wncyss}%
\renewcommand\encodingdefault{OT2}%
\normalfont
\selectfont}
\DeclareTextFontCommand{\textcyr}{\cyr}

\def\be{\begin{equation}}
\def\ee{\end{equation}}
\def\ba{\begin{eqnarray}}
\def\ea{\end{eqnarray}}
\def\bs{\begin{subequations}}
\def\es{\end{subequations}}
\def\a{\alpha}

\def\de{\delta}

\def\e{\epsilon}

\def\t{\tau}  
\def\s{\sigma}

\def\N{\nabla}

\def\cK{\mathcal{K}}

\def\cP{\mathcal{P}}

\def\cS{\mathcal{S}}

\def\cV{\mathcal{V}}

\def\ds{d_{\rm S}}
\def\dh{d_{\rm H}}

\def\p{\partial}

\def\B{\Box}
\newcommand{\Eq}[1]{(\ref{#1})}
\def\com{\color{magenta}}
\def\cob{\color{blue}}
\newcommand{\au}[2]{#2 #1}
\newcommand{\aud}[3]{#3 #1 #2}
\newcommand{\aut}[4]{#4 #1 #2 #3}
\newcommand{\oarX}[1]{\href{http://arxiv.org/abs/#1}{(arXiv:{\com #1})}}
\newcommand{\arX}[1]{\href{http://arxiv.org/abs/#1}{(arXiv:{\com #1})}}
\newcommand{\doin}[6]{\href{http://dx.doi.org/#1}{\cob \textit{#2} #3 \textbf{#4} #5}}

\newcommand{\doinn}[5]{\href{http://dx.doi.org/#1}{\cob \textit{#2} \textbf{#3} #4}}

\newcommand{\doij}[5]{\href{http://dx.doi.org/#1}{\cob \textit{#2} \textbf{#3} #4}}
\newcommand{\tia}[1]{#1}
\newcommand{\books}[4]{#4 \emph{#1} (#3: #2)}

\begin{document}

\date{October 17, 2013}

\begin{center}
\doinn{10.1088/1751-8113/47/35/355402}{J.\ Phys.\ A: Math.\ Theor.}{47}{(2014) 355402}{2014} \arX{1310.4957} \hfill October 17, 2013
\end{center}

\title{Nonlocality in string theory}

\author{Gianluca Calcagni}
\address{Instituto de Estructura de la Materia, CSIC, calle Serrano 121, 28006 Madrid, Spain}
\author{Leonardo Modesto}
\address{Department of Physics \& Center for Field Theory and Particle Physics, Fudan University, 200433 Shanghai, Peoples's Republic of China}
\eads{\mailto{calcagni@iem.cfmac.csic.es}, \mailto{lmodesto@fudan.edu.cn}}

\begin{abstract}
We discuss an aspect of string theory which has been tackled from many different perspectives, but incompletely: the role of nonlocality in the theory and its relation to the geometric shape of the string. In particular, we will describe in quantitative terms how one can zoom out from an extended object such as the string in such a way that, at sufficiently large scales, it appears structureless. Since there are no free parameters in free-string theory, the notion of large scales will be unambiguously determined. In other words, we will be able to answer the question: How and at which scale can the string be seen as a particle? In doing so, we will employ the concept of spectral dimension in a new way with respect to its usual applications in quantum gravity. The operational notions of worldsheet and target spacetime dimension in string theory are also clarified and found to be in mutual agreement.
\end{abstract}


\hspace{1.6cm} {\footnotesize Keywords: string theory, nonlocal theories and models, string field theory}


\maketitle


\section{Introduction}

String theory is perhaps the best studied candidate theory for physics beyond the Standard Model of particles. Still, its mathematical richness is far from being fully tapped. In this work, we examine the relation between nonlocality and extended objects, in particular the string. Depending on the context in which it arises (free or interacting string), nonlocality affects the way in which an observer would perceive both a propagating string and the dimensionality of spacetime.

In systems defined in continuous spacetimes, \emph{nonlocality} is the appearance of an infinite number of derivatives in the kinetic terms $\phi f(\p)\phi$ of fields or, what is equivalent \cite{PU,BOR}, by interactions $\int\rmd x' \phi(x) F(x-x')\phi(x')$ dependent on noncoincident points. String theory is not formulated on spacetime but its very nature makes it nonlocal. Imagining strings to evolve in a predetermined spacetime can help to intuitively understand why. When strings propagate, their trajectories are not one-dimensional worldlines but two-dimensional worldsheets, which are strips for open strings and tubular surfaces for closed strings. When strings meet, their interactions are not pointwise but extended to a finite region of space; this feature is at the core of the ultraviolet (UV) finiteness that the theory is believed to have. The string is nonlocal also in the quantitative sense of its field-theory limit \cite{EW89b}, since the particle-field content of the string spectrum evolves according to the nonlocal operators
\be\label{fp}
f(\B)\sim \B\rme^{-c\a'\B}\,,
\ee
where $\B=\p_\mu\p^\mu$ is the Laplace--Beltrami operator, $\mu$ runs from 0 to $D$, $c=O(1)$ is a constant, and $\a'$ is the Regge slope. This form factor arises in string field theory (SFT \cite{ohm01,FK}), from the way in which open strings interact; after a field redefinition, the nonlocality in interaction terms is transferred to the kinetic terms which then take the form of \Eq{fp}. The closed-string case gives the same $f$ but with $c\to c/2$. In turn, the solutions of string-field equations are given by the superposition of a one-parameter family of states in a conformal field theory \cite{FK}. These `surface states' obey a universal diffusion-like equation where diffusion time is the parameter of the family and the spatial generator (i.e., what would be the Laplacian in ordinary space) is a combination of Virasoro operators. This structure filters from the conformal-field-theory formulation down to the spacetime limit of the system to give the exponential form factor \Eq{fp} \cite{cuta7}.

In general, nonlocal operators are troublesome. On one hand, the Cauchy problem can be ill defined or highly nonstandard \cite{PU,Pau53}, since it would entail an infinite number of initial conditions $\phi(0)$, $\dot\phi(0)$, $\ddot\phi(0)$, \dots, representing an infinite number of degrees of freedom. As the Taylor expansion of $\phi(t)$ around $t=0$ is precisely given by the full set of initial conditions, specifying the Cauchy problem would be tantamount to knowing the solution itself \cite{MoZ}. This makes it very difficult to find analytic solutions to the equations of motion. On the other hand, a particle interpretation of a quantum field theory may even be absent, due to the replacement of poles in the propagators with branch cuts \cite{BOR,doA92}. Moreover, causality may be violated at microscopic scales \cite{PU}. However, the type of nonlocality \Eq{fp} found in SFT (or with form factors $\exp{H(\Box)}$, where $H(\Box)$ is an entire function \cite{Tom97}) is far more benign than it was earlier deemed \cite{EW89b}. The entire function $\rme^\B$ introduces neither extra poles (as in ghost-plagued theories with derivatives of higher but finite order \cite{EW89b}) nor branch cuts \cite{Efi77}. Also, the Cauchy problem can be reformulated in such a way that the infinite number of degrees of freedom, so characteristic of nonlocal theories, is absorbed into suitable field redefinitions via perturbative \cite{BK1} and nonperturbative techniques \cite{cuta3}. Causality, too, is respected in theories with exponential operators, both at the microscopic and at the macroscopic level \cite{Tom97,Efi77, Modesto}. All this is somewhat reassuring, as the same conformal structure leading to equation \Eq{fp} is responsible for the UV finiteness of string theory.

Nonlocal theories were recognized early as describing physical entities which are not pointwise \cite{Pau53}. An example is nonlocal electrodynamics, where the electron becomes a spherical charge distribution of finite radius \cite{Efi77,NaD}. Similarly, the exponential form factors appearing in a class of super-renormalizable theories of gravity \cite{Modesto,BGKM} may be conjectured to be a manifestation of underlying extended objects. In general, `nonlocality' is tightly related to the appearance of `extended objects', but a precise connection between these two concepts is still lacking. It is our purpose to clarify such an issue both in nonlocal theories and, especially, for the string case, giving it a precise physical meaning in terms of the resolution of the string worldsheet.

Instead of starting from SFT and the form factor \Eq{fp}, it is more instructive to discard interactions and consider the \emph{free} string. In 1990, Kato showed that a quantized particle with a certain nonlocal kinetic term reproduces the spectrum of open bosonic free string and the Veneziano amplitude \cite{Kat90}. Within the same model (reviewed and clarified in section \ref{sec2}), we will show that the trajectory of the classical particle is two-dimensional in the UV, that is, a worldsheet (section \ref{sec3}). Thus, the physical meaning of the relation between Kato's model and open bosonic string theory is improved, but we will also be able to comment on the closed-string case and on strings attached to branes (section \ref{sec4}).


\section{Effective nonlocal action}\label{sec2}

All the information we need is in the bosonic sector, so we ignore supersymmetry. The Polyakov action for the open bosonic string is (see, e.g., \cite{Zwi09}; here $\a'=1$)
\be
\label{poly}
\cS=\frac{1}{2\pi}\int_{-\infty}^{+\infty}  \! \rmd\t  \! \int_0^\pi\rmd\s\,\left(\p_\t X_\mu\p_\t X^\mu-\p_\s X_\mu\p_\s X^\mu\right),
\ee
where $\tau$ and $\s$ are coordinates parametrizing, respectively, the (infinite) length and the (finite) width of the worldsheet, $X^\mu(\t,\s)$ are $D=26$ scalars, and Greek indices are contracted with the Minkowski metric $\eta^{\mu\nu}$. We gauge fixed the worldsheet metric to the flat one by combining worldsheet diffeomorphisms and a Weyl transformation. The equation of motion proceeding from \Eq{poly} is
\be\label{steom}
(-\p_\t^2 +\p_\s^2) X^\mu=0\,.
\ee
To solve it, one should specify a set of boundary conditions. For the free open string, these would be the Neumann boundary conditions $\p_\s X^\mu(\t,0)=0=\p_\s X^\mu(\t,\pi)$. Plugging the solution back into the action \Eq{poly}, one would trivially get $\cS=b$, where $b$ is a constant (not necessarily finite) boundary term. In fact, for any solution $X_{\rm sol}$ of \Eq{steom} (integration domains omitted),
\ba
\fl 2\pi\cS[X_{\rm sol}] &=& \int\rmd\s\,\eta_{\mu\nu}X^\mu_{\rm sol}\p_\t X^\nu_{\rm sol}\Big|_{\t=-\infty}^{\t=+\infty}-\int\rmd\t\,\eta_{\mu\nu}X^\mu_{\rm sol}\p_\s X^\nu_{\rm sol}\Big|_{\s=0}^{\s=\pi}\nonumber\\
&&-\int\rmd\t  \! \int\rmd\s\,\eta_{\mu\nu}X^\mu_{\rm sol}\left(\p_\t^2-\p_\s^2\right)X^\nu_{\rm sol}\nonumber\\
&=& 2\pi b-\int\rmd\t\,\eta_{\mu\nu}X^\mu_{\rm sol}\,\p_\s X^\nu_{\rm sol}\Big|_{\s=\pi}+\int\rmd\t\,\eta_{\mu\nu}X^\mu_{\rm sol}\,\p_\s X^\nu_{\rm sol}\Big|_{\s=0},\label{bounda}
\ea
and the last two terms vanish for the Neumann boundary conditions. However, we want to obtain a \emph{nontrivial} effective action for the string in the limit where its length can be ignored, so that one can treat it as a particle. To perform this zooming out correctly, one should integrate the degrees of freedom along the string length, but this is not possible if we specify the Neumann conditions on both endpoints of the string. Therefore, we must define one of the endpoints by a generic Dirichlet boundary condition\footnote{As one can easily check, the final result \Eq{acf2} will not depend on the choice of the endpoint, thanks to the reparametrization invariance $\s\to\pi-\s$.}, plug the corresponding solution $\bar X$ back into equation \Eq{poly}, and integrate over the length parameter $\sigma$. The resulting effective action $\cS[\bar X]$ will encode the degrees of freedom we are looking for. They will correspond to a `particle', but not in the usual sense of a pointwise object with local dynamics.

Let us see this procedure in detail. We Fourier transform $X$ in the worldsheet coordinate $\tau$:
\ba
\fl X^\mu(\t,\s) &=& \int_{-\infty}^{+\infty}\frac{\rmd k}{2\pi}\,\rme^{\rmi k\t}\,g(k,\s)\,x^\mu(k)\,,\qquad x^\mu(k):=\int_{-\infty}^{+\infty}\rmd\t'\,\rme^{-\rmi k\t'}x^\mu(\t')\,,\label{sol0}\\
\fl x^\mu(\t)   &:=&  X^\mu(\t,0)\,,\label{xX}
\ea
where we singled out the Fourier transform $x(k)$ of the $\s=0$ endpoint and $g(k,\s)$ is some kernel function determined by the boundary conditions. According to equation \Eq{xX} and imposing only $\p_\s X^\mu(\t,\pi)=0$,
\be\label{cong}
g(k,0)=1\,,\qquad \p_\s g(k,\s)\Big|_{\s=\pi}=0\,. 
\ee
Applying equation \Eq{steom} to \Eq{sol0}, we get $(\p_\s^2+k^2)g(k,\s)=0$, the harmonic oscillator with solution $g(k,\s)=b_k\sin(k\s)+c_k\cos(k\s)$. The conditions \Eq{cong} require, respectively, that $b_k=1$ and $c_k=\sin(k\pi)/\cos(k\pi)$. The solution of \Eq{steom} is thus
\ba
\bar X^\mu(\t,\s) &=& \int_{-\infty}^{+\infty}\frac{\rmd k}{2\pi}\,\rme^{\rmi k\t}\frac{\cos[k(\pi-\s)]}{\cos (k\pi)}\,x^\mu(k)\,,\label{Nsol}\\
\bar X^\mu(\t,0)  &=&  x^\mu(\t)\,.
\ea
Notice that the presence of the denominator $\cos(k\pi)$ does not necessarily imply that \Eq{Nsol} is ill defined for semi-integer $k$, since the $k$ dependence of the rest of the integrand is unspecified. If we further imposed $\p_\s X^\mu(\t,0)=0$, we would get $c_k=0$ and $k$ would take only integer values. One would then obtain the usual free-string solution 
\be\label{neu}
X_{\rm Neu}^\mu=p^\mu\tau+\sum_{n\in\mathbb{Z}}\rme^{\rmi n\t}\cos(n\s)\,x^\mu(n)\,,
\ee
which coincides with \Eq{Nsol} when $k=n$ (the integral is replaced by a sum) and up to the linear term $p^\mu\tau$ (it has been added in \Eq{neu} as a part of the general solution, but it does not play any role in \Eq{Nsol} and in what follows). 

However, momenta $k$ are continuous since the function $x(\t)$ is not a Cauchy datum. Equation \Eq{Nsol} represents a string with the $\s=\pi$ endpoint free, not bound to any brane, and the other one unspecified. To give the solution \Eq{Nsol} a physical meaning, we plug it into the action \Eq{poly} and integrate over $\s$. One thus obtains an effective action for the endpoint coordinate $x^\mu(\t)$, which is then interpreted as a worldline whose dynamics is characterized by a nonlocal form factor (integration domains are omitted from now on):
\ba
\fl S &:=& \cS[\bar X]-b \stackrel{{\textrm{\tiny \Eq{bounda}}}}{=} \frac{1}{2\pi}\int\rmd\t\, \bar X_\mu\p_\s \bar X^\mu\Big|_{\s=0}= \frac{1}{2\pi}\int\rmd\t\, x_\mu(\tau) \int\frac{\rmd k}{2\pi}\,\rme^{\rmi k\t}k\,\tan(k\pi)\,x^\mu(k)\nonumber\\
 \fl &\,=& -\frac{1}{2}\int\rmd\t\, x_\mu(\tau) \int\frac{\rmd k}{2\pi}\,\rme^{\rmi k\t}\cK(\rmi k)\,x^\mu(k)= -\frac{1}{2}\int\rmd\t\, x_\mu(\tau) \cK(\p_\t)\int\frac{\rmd k}{2\pi}\,\rme^{\rmi k\t}x^\mu(k)\nonumber\\
\fl  &\,=& -\frac{1}{2}\int\rmd\t\, x_\mu(\t)\, \cK(\p_\t)\, x^\mu(\t)\,,\label{acf2}
\ea
where
\be\label{ck}
\cK(\rmi k)=-\frac{1}{\pi}\,\tan(\pi k)\,k\,,\qquad \cK(\p_\t)=\frac{1}{\pi}\tanh(\pi\p_\t)\p_\t\,.
\ee
For small momenta $k$, $\cK(\rmi k)\simeq -k^2\to \p_\t^2$ and the action \Eq{acf2} becomes that for a nonrelativistic point particle:
\be\label{popa}
S\simeq S_{\rm p}=-\frac{1}{2}\int\rmd\t\, x_\mu(\t) \p_\t^2 x^\mu(\t)\,.
\ee
Notice that this can also be regarded as a relativistic particle in the gauge of unit $D$-velocity, which can always be chosen by a worldline reparametrization \cite{Kat90}. As the Fourier transform of the Hamiltonian $H\propto \dot x^2$ from equation \Eq{popa} is proportional to $k^2$, the point particle can be regarded as the low-energy limit of the string.

Even if the solution \Eq{Nsol} is represented as an integral over continuous momenta $k$, the system possesses the conformal symmetry of the free string. Classically, \Eq{acf2} is required to be invariant under infinitesimal reparametrizations $x(\t)\to x'(\t)=x(\t+\de\t)$, where $\de\t=\e\,\rme^{\rmi k \t}+{\rm c.c.}$ (c.c.\ stands for complex conjugate) and $\e$ is small. This implies that, if $x^\mu(\t)$ is solution to the equation of motion $\cK(\p_\t)x^\mu=0$, then also ${x'}^\mu=x^\mu+\de x^\mu=x^\mu+\p_\t x^\mu \de\t$ is a solution, implying the constraint \cite{CHL}
\be
\cK(\p_\t)\de x^\mu=0 \qquad\Rightarrow\qquad \cK(\p_\t+\rmi k)\p_\t x^\mu=0\,,
\ee
which is satisfied by equation \Eq{ck} only if $k=n\in\mathbb{Z}$ is an integer (obviously, this range avoids poles in the $g$ function in \Eq{Nsol}). By Noether's theorem, the conserved charges associated with this symmetry are \cite{Kat90,CHL}
\be
L_n=\frac12\int_{-\pi}^\pi\rmd\tau\,\rme^{\rmi n\t}p_\mu(\t)\,p^\mu(\t)\,,
\ee
where $p^\mu:=\de S/\de(\p_\t x^\mu)$ is the conjugate momentum. The $L_n$ obey the Witt algebra $\{L_m,L_n\}=(m-n)L_{m+n}$, i.e., a Virasoro algebra with vanishing central charge. 

The infinite countable number of modes of the string are all contained in the operator $\cK$. Let us define the nonlocal operator $f_0(\p_\t)$ (using the notation of \cite{Kat90}) such that $\cK=\p_\t(f_0\p_\t\,\cdot)$. After integrating by parts, the action \Eq{acf2} is rewritten as
\be\label{acff}
S=\frac12\int\rmd\t\,(\p_\t x_\mu)\,f_0(\p_\t)\, \p_\t x^\mu\,.
\ee
The inverse operator $1/f_0$ admits the series representation
\be\label{sef}
\frac{1}{f_0(\p_\t)}=1+2\sum_{n=1}^{+\infty}\frac{\p_\t^2}{\p_\t^2+n^2}\,,
\ee
which can be used to recast $x$ as an infinite superposition of modes:
\be\label{modes}
x^\mu(\t)=x_0^\mu(\t)+2\sum_{n=1}^{+\infty}x_n^\mu(\t)\,,
\ee
where
\be
x_0^\mu:=f_0(\p_\t) x^\mu\,,\qquad x_n^\mu:=\frac{\p_\t^2}{\p_\t^2+n^2}\,f_0(\p_\t) x^\mu\,.
\ee
Up to a total derivative, the action \Eq{acff} reads \cite{PU,Kat90}
\be
 S=\frac12\int\rmd\t\,\eta_{\mu\nu}\left[\p_\t x_0^\mu\p_\t x_0^\nu+2\sum_{n=1}^{+\infty}\left(\p_\t x_n^\mu\p_\t x_n^\nu-n^2x_n^\mu x_n^\nu\right)\right].
\ee

Upon quantization, the model \Eq{acf2} turns out to be unitary with positive semidefinite Hamiltonian $\hat H=\hat L_0$. The conserved charges associated with the reparametrization symmetry are the $\hat L_n=(1/2)\eta_{\mu\nu}\sum_{m\in\mathbb{Z}}y_{n-m}^\mu y_m^\nu$, where $y_0^\mu=p^\mu$ and $y_n^\mu=\a_n^\mu$ are the annihilation operators into which $x^\mu_n$ is decomposed \cite{Kat90,CHL}. The generators $\hat L_n$ obey the Virasoro algebra $[\hat L_m,\hat L_n]=(m-n)\hat L_{m+n}-(c/12)m(m^2-1)\de_{m+n,0}$. Therefore, the infinite number of degrees of freedom encoded in the nonlocal action \Eq{acf2} are nothing but the open bosonic string spectrum. The nonlocal model realizes the same Fock space of physical states of the full theory.

To summarize so far, the action \Eq{poly} gives rise to a nonlocal model for a particle when considering the system at a lower resolution, i.e., when looking at the string sufficiently far away to lose the perception of it as an extended object. In the next section, we will render the zooming-out resolving procedure and the meaning of `sufficiently far away' quantitative. 


\section{Spectral dimension}\label{sec3}

Equation \Eq{acf2} represents a particle but with a peculiar type of dynamics, such that it cannot be really treated as a `point'. The construction of section \ref{sec2} has been carried out in \cite{Kat90,CHL}, while here we have improved its physical interpretation. We further do so in the following, showing how the action \Eq{acf2} interpolates between a point particle in the infrared (IR) to a worldsheet in the UV. The transition will be uniquely determined by the string scale.


\subsection{Spectral dimension of the string}

To get acquainted with the necessary tools, it is instructive to have a short digression. To probe the geometry of a set, one can place a test particle on it and let it diffuse from some initial point. The diffusion clock is a length scale $\ell$, whose inverse represents the resolution at which the process is studied. This interpretation is an improvement over the usual declaration that $\ell$ is an `abstract diffusion time', which has no clear meaning when the set to probe is spacetime itself. Thus, the diffusion equation is rather regarded as a running equation (beta function). One typical context in which the diffusion approach is applied is quantum gravity, where spacetime geometry is deformed by quantum effects. In this case, diffusion is often governed by an equation of the form $[\p/\p\ell^2+\cK(\p_x)]P(x,x';\ell)=0$, where $P$ is the probability density function of the process with initial point $x'$, and the integro-differential operator $\cK(\p_x)$ is determined by the effective dynamics of fields. In the absence of quantum-gravity effects, $-\cK$ is the Laplacian $\N_x^2$. Moreover, diffusion on space and time is well defined only if the time direction is Wick rotated to a Euclidean configuration, lest the diffusion equation be parabolic.

Here, unlike in standard applications of the diffusion equation in quantum gravity, we want to probe not the geometry of spacetime but \emph{the trajectory of the probe itself}. As we were unable to find instances in the literature where the diffusion equation was employed in this way, we assume the procedure to be unfamiliar to the general reader and take the ordinary particle \Eq{popa} as an example. This is a classical-mechanics system parametrized by one worldline dimensionless parameter $\t$ and with kinetic operator $-\p_\t^2$. After carrying out the Wick rotation $\t\to-\rmi\t$, one defines the diffusion equation on the worldline:
\be\label{dif1}
\hspace{-0.2cm} \left(\frac{\p}{\p L^2}-\p_\t^2\right)P[\t',\t;\ell(L)]=0\,, \qquad 
P(\t,\t',0)=\de(\t-\t'),
\ee
where $L:=\ell/\ell_*$, $\ell_*$ is a fixed reference length scale, and the initial condition at $\ell=0$ states that the probe is pointwise (i.e., it can detect individual points) when the resolution $1/\ell$ is infinite. The solution (normalized to 1) is a one-dimensional Gaussian, $P[\t',\t;\ell(L)]=\exp[-(\t-\t')^2/(4L^2)]/\sqrt{4\pi L^2}$. When $\t'= \t$, integrating in time and dividing by the divergent integral $\int\rmd\t$, one obtains the return probability $\cP(\ell):=\int\rmd\t\,P(\t,\t;\ell)/(\int\rmd\t)=P(\t,\t;\ell)\propto \ell^{-1}$, from which the \emph{spectral dimension} descends:
\be\label{ds}
\ds(\ell):=-\frac{\rmd\ln\cP(\ell)}{\rmd\ln\ell}\,.
\ee
In this example, $\ds=1$ at all scales. Its interpretation is straightforward and is one of the central tenets of the present paper: \emph{In classical mechanics, $\ds$ is the dimension of the trajectory of the particle}. Here we have a standard point particle spanning a worldline, which is a one-dimensional set.

The result $\ds=1$ is valid only at the classical level, where the particle trajectory is smooth. The path of a quantum particle (in a classical spacetime) is so irregular that it is nowhere differentiable \cite{FeH} and its Hausdorff dimension is $\dh=2$ \cite{AbW}. Here we calculated the spectral rather than the Hausdorff dimension, but $\ds>1$ is expected for a quantum path anyway.

\begin{figure}
\centering
\includegraphics[width=9cm]{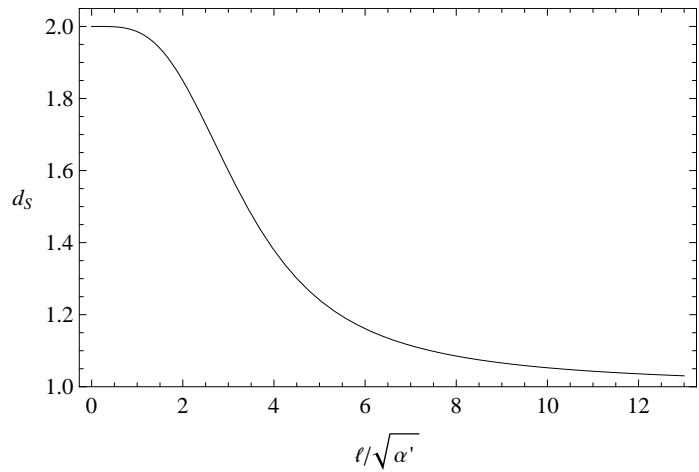}
\caption{\label{fig1} The dimension of the path of the classical nonlocal particle \Eq{acf2}. At UV scales $\ell\lesssim 10\sqrt{\a'}$, the structure of the string worldsheet becomes apparent.}
\end{figure}

Let us now move to the nonlocal action \Eq{acf2}. The diffusion equation \Eq{dif1} is replaced by
\be\label{dif2}
\left[\frac{\p}{\p L^2}+\cK(\rmi\p_\t)\right] P=0\,,\qquad P(\t,\t';0)=\de(\t-\t'),
\ee
whose solution is
\be
P(\t,\t';\ell)=\int \frac{\rmd k}{2\pi}\,\,\rme^{-(\ell/\ell_*)^2\cK(k)}\rme^{\rmi k(\t-\t')},
\ee
where $\cK(k)=\tanh(\pi k)\,k/\pi$. Notice that $P>0$ and the diffusive process is well defined (i.e., it admits a probabilistic interpretation). In the IR limit $\ell\to+\infty$, the exponential is dominated by small momenta $k$ where $\cK(k)\simeq k^2\ll 1$, so one recovers the local particle case. Recalling that this model stems from a zooming out from the free string, the only possible choice for the scale $\ell_*$ is $\sqrt{\a'}$, which we set to 1 at the beginning. The spectral dimension \Eq{ds} is then
\be\label{dsel}
\ds(\ell)=2\frac{(\ell/\ell_*)^2\int \rmd k\,\rme^{-(\ell/\ell_*)^2\cK(k)}\cK(k)}{\int \rmd k\,\rme^{-(\ell/\ell_*)^2\cK(k)}}\,,
\ee
and it is shown in figure \ref{fig1}. In the UV, $\ds$ tends to 2. By itself, this would not be sufficient for concluding that the string worldsheet structure is recovered at small scales. In fact, there exist sets with integer dimension which are nonsmooth (e.g., the boundary of the Mandelbrot set), while nonplanar geometries with $\ds=2$ occur in various quantum-gravity models. However, we also know \cite{Kat90,CHL} that the model \Eq{acf2} has a Virasoro algebra and conformal symmetry, characteristics that no two-dimensional set except the worldsheet would possess. This concludes the proof of the main claim.

The profile of the spectral dimension provides a quantitative tool for zooming in to and out from the string. At scales $\ell\lesssim 10\sqrt{\a'}$, the extended nature of the string unfolds. 

We can give an intuitive explanation of why we retained continuous momenta instead of discrete ones in expressions such as \Eq{Nsol} and \Eq{dsel}. The ordinary free string is endowed with a countable number of modes following from conformal invariance; this invariance is manifest in the UV. On the other hand, the effective theory \Eq{acf2} is a limit of the full theory such that the infinite countable number of modes has been swept into a nonlocal operator (equation \Eq{sef}) and, at the same time, there exists an IR limit where conformal invariance is lost in the continuum of the point-particle reparametrization symmetry\footnote{Another way to state this difference between full and effective theory is the following. We have seen that the Polyakov action \Eq{poly} is trivial on the solution \Eq{neu} with Neumann boundary conditions at both endpoints, $\cS[X_{\rm Neu}]=b$. For Lagrangian systems, an expansion of the action around a field solution $\phi_0$ yields $\cS[\phi_0+\delta\phi]=\cS[\phi_0]+\delta^2 \cS/\delta\phi^2|_{\phi=\phi_0}(\phi-\phi_0)^2/2+\dots$, where the first-order term vanishes by virtue of the equations of motion $\de\cS/\delta\phi|_{\phi=\phi_0}=0$. Inserting this in the path integral $\rme^{\rmi \cS}$ and ignoring the first contribution as a normalization, one interprets higher-order terms as a loop expansion. In systems such as the free string where $\cS[\phi_0]$ is trivial, this expansion is purely quantum inasmuch as there are no surviving classical degrees of freedom, while the leading nontrivial term in the series is already at one-loop level. On the other hand, in the effective theory \Eq{acf2} the term $\cS[\phi_0]$ is nontrivial and contains residual degrees of freedom.}. In other words, the integral representation with continuous $k$ allowed for the possibility of quantitatively realizing the zooming in to and out from the string. This can be checked by replacing sums instead of integrals in the above expressions (in particular, \Eq{dsel}) and recalculating $\ds$ numerically: while the UV limit $\ds\simeq 2$ would be the same, the IR point-particle limit $\ds\simeq 1$ would not be recovered.


\subsection{Spectral dimension of spacetime}

The effective dimensionality of the string as encoded by equation \Eq{ds} has nothing to do with the various notions of \emph{spacetime} dimension in string theory. The topological dimension $D$ of the target spacetime is simply the number of scalars $X^\mu$, determined by the cancellation of the conformal anomaly. The spectral dimension of the target spacetime can be derived from the SFT form factor \Eq{fp} and is more in line with traditional computations of $\ds$ where the probe is a field and the set to inspect is spacetime. The standard procedure is to consider the diffusion equation with the Laplacian replaced by the Wick-rotated form factor of the given field theory, in this case \Eq{fp}. The fact that this kinetic operator arises after a field redefinition of a system with ordinary kinetic term and interactions dressed with nonlocal operators should not be of concern. The diffusion equation method is essentially based on the free-field limit of the theory on the assumption that interactions do not contain derivative terms. In effective bosonic SFT, this is achieved precisely after the above-mentioned field redefinition, when interactions are made local and all nonlocality is transferred to the kinetic term. Keeping nonlocal operators in the interaction terms would not account for all the dynamical effects of the system on the diffusion of the test probe.

\begin{figure}
\centering
\includegraphics[width=9cm]{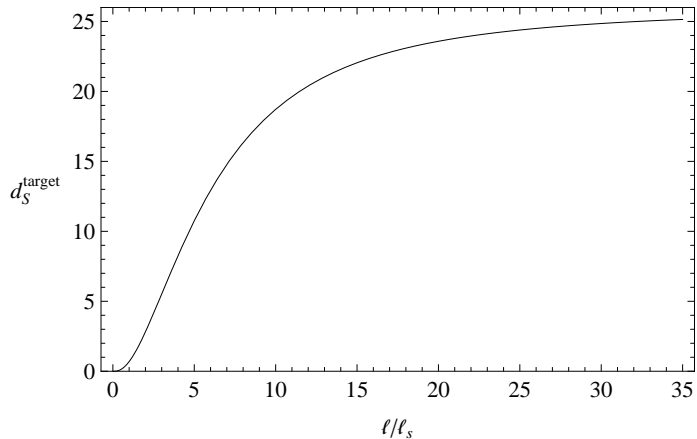}
\caption{\label{fig2} Spectral dimension of the target spacetime in string field theory. In the IR limit, the spectral and topological dimension coincide ($D=26$ in this example), while in the UV limit the spectral dimension vanishes.}
\end{figure}

Using equation \Eq{fp} and imposing the diffusion equation (with a pointwise initial condition)
\be
\fl\left[\frac{\p}{\p\ell^2}-f(\B_{\rm E})\right]P_{{\rm target}}(x,x';\ell)=0\,,\qquad P_{{\rm target}}(x,x';0)=\de(x-x')\,,
\ee
where $\B_{\rm E}=\N^2_x$ is Wick-rotated, the solution as a momentum-space integral is
\be
P_{{\rm target}}(x,x';\ell)=\int \frac{\rmd^D p}{(2\pi)^D}\,\,\exp\left[-\ell^2 p^2\rme^{\ell_{\rm s}^2p^2}\right]\rme^{\rmi p\cdot(x-x')},
\ee
where $\ell_{\rm s}:=\sqrt{c\a'}$. The return probability is
\ba
\cP_{{\rm target}}(\ell) &=& P_{{\rm target}}(x,x;\ell)=\Omega\int_0^{+\infty} \rmd p\,p^{D-1}\exp\left[-\ell^2 p^2\rme^{\ell_{\rm s}^2p^2}\right]\nonumber\\
&=& \ell^{-D}\Omega\int_0^{+\infty} \rmd q\,q^{D-1}\exp\left[-q^2\rme^{(\ell_{\rm s}/\ell)^2q^2}\right]\,,\label{cptar}
\ea
where we ignored a diverging volume factor $\cV=\int\rmd^Dx$, $\Omega$ is a constant coming from the integration of the solid angle, $p=|p|$ and $q=\ell p$ is dimensionless. The spectral dimension of spacetime is then
\ba
\ds^{\,{\rm target}}(\ell)&=&-\frac{\rmd\ln \cP_{{\rm target}}(\ell)}{\rmd\ln\ell}\nonumber\\
&=&2\frac{\int_0^{+\infty} \rmd q\,q^{D+1}\exp\left\{-q^2[(\ell_{\rm s}/\ell)^2+\rme^{(\ell_{\rm s}/\ell)^2q^2}]\right\}}{\int_0^{+\infty} \rmd q\,q^{D-1}\exp\left[-q^2\rme^{(\ell_{\rm s}/\ell)^2q^2}\right]}\,.
\ea
In the IR limit $\ell/\ell_{\rm s}\gg 1$, the spectral dimension coincides with the topological dimension of the target spacetime, $\ds^{\,{\rm target}}\simeq D$; this can be seen directly from equation \Eq{cptar}, as the integral becomes independent of $\ell$. In the UV, a numerical integration shows that (figure \ref{fig2})
\be\label{sftds}
\ds^{\,{\rm target}}\simeq 0\qquad {\rm (UV)}\,.
\ee

The crucial point is that $\rme^{\B^n}$ nonlocalities, when appearing in kinetic terms of field theories, erase the information on the structure of the probe. The operator $\rme^\B$ naively gives rise to an extended object when applied to the initial condition, since $\rho_l(x-x')=\rme^{l^2\B}\de(x-x')$ is a Gaussian of width $l$ \cite{SSN}, but this `fattening' of the source is completely negligible inside the exponential well created by the form factor, as one can see from the momentum kernel $\exp[-p^2(\ell^2\rme^{\ell_{\rm s}p^2}+l^2)]$ in the limit of large $p$. Thus, field theories with exponential operators (including SFT) can only see point particles by construction (consistently, the microcausality lightcone has a singular vertex \cite{Efi77}), and equation \Eq{sftds} holds also for an extended source. However, in the case of string theory we have extra classical-mechanics information which allowed us to find the dimension of the probe itself rather than of its environment. Although at high energy scales the spectral dimension \Eq{sftds} of spacetime vanishes, this result is insensitive of the intrinsic size of the probe, which we saw to be actually (after Euclideanization) two-dimensional at energies comparable with the string tension (figure \ref{fig1}). Having $\ds=0$ with a source with undetermined structure means that the probe, whatever its details, does not diffuse in spacetime because it `fills' spacetime completely. Therefore, a nondiffusing extended probe signals a spacetime of the same dimensionality, in this case 2.



\section{Generalizations and discussion}\label{sec4}

The above results admit various generalizations. 
\begin{itemize}
\item\emph{Closed string}. The effective nonlocal action is made up of two copies of equation \Eq{acf2} (representing the left-moving and right-moving sectors) plus a constraint linear in the coordinates and identifying the zero modes of the copies \cite{CHL}. Thus, the running of the spectral dimension is the same as for the open string, and the role of topology is negligible due to zooming-out effects. However, it is predominant in \cite{AtW}, where it was shown that, at high temperature, the thermodynamical dimension of the target spacetime is $D_T:=\ln(F/V)/\ln T\simeq 2$ for closed strings and $\simeq 1$ for open strings \cite{AtW}, where $F$, $V$, and $T$ are, respectively, the free energy, volume, and temperature of the system. The thermodynamical and spectral dimensions have been conjectured to be the same quantity for fractals \cite{Akk2}, so one might expect that $D_T\sim\ds^{\,{\rm target}}$. However, both the speculative nature of this equivalence and the resolution limitations of the present approach prevent us from making a detailed comparison with the findings of \cite{AtW}. 
\item\emph{Dirichlet boundary conditions}. This case is more tricky, as it amounts to having at least one endpoint of the open string stuck at a D-brane. For instance, imposing the Dirichlet condition $X^\mu(\t,\pi)=0$ on one endpoint, one gets $\cK(\rmi k)=-\cot(\pi k)k/\pi\to\coth(\pi\p_\t)\p_\t/\pi$. However, the small-momentum limit yields a constant term plus a Laplacian with wrong sign, $\cK(\rmi k)\simeq -1/\pi^2+k^2/3$, which signals a nonstandard local limit. After carrying out Wick rotation, one can compute the spectral dimension as before, obtaining a running from 2 in the UV to infinity in the IR. This behaviour has no obvious diffusion interpretation, but understandably so: Since the string is stuck at one endpoint, when zooming out from it one would see a still particle. 
\item\emph{Kato's family}. In \cite{Kat90}, the operator \Eq{ck} was the $s=0$ element of the one-parameter family $\cK_s(\rmi k)=-(k/\pi)\sin(\pi k)/\cos[(\pi-s)k]$. It is easy to check that the spectral dimension of these models runs from 0 in the UV to 1 in the IR. When $s\neq 0$ is sufficiently small, there is an intermediate peak larger than 1, which increases to the value 2 and is pushed towards $\ell=0$ as $s$ decreases to zero. The nonlocal action \Eq{acf2} with equation \Eq{ck} is the only one associated with a worldsheet geometry in the UV, as expected by the extreme rigidity of the conformal structure of string theory.
\end{itemize}

To summarize, via an explicit example we sharpened the relation between the extended scale-dependent nature of the fundamental degrees of freedom of nonlocal systems and the value of the spectral dimension. It is not a new conjecture that the spectral dimension is a viable tool to use for understanding the meaning of form factors in nonlocal field theories, nor is it a new conjecture that the latter are associated, in general, with extended objects. In this paper, we presented a positive test result for both conjectures. In the example under inspection, we know that the theory at high energy is described by strings, while at low energy it is described by point particles. Therefore, the effective dimension of the (trajectory of the) fundamental degree of freedom is 2 at high energy and 1 at low energy. Although this sounds fairly obvious from any simple cartoon of the string when zoomed in to and out from, here we applied a novel method to make this dimensional flow under analytic control at any scale. 

The first advantage is having a clear-cut notion of what we mean by `dimensionality of the degrees of freedom', scale by scale; we achieved this through a resolution-changing probing of the string or, loosely speaking, by a `probing of the probe'. The interpolating theory is defined by a nonlocal action with a specific form factor. The second advantage is that, thanks to the inspection of the string-related nonlocal theory via a resolution-dependent tool, we have given more physical insight to the effective nonlocal model \Eq{acf2} \cite{Kat90}, which was lacking a concrete geometric interpretation. The third and last advantage is the clarification of how the more traditional notion of spectral dimension of spacetime (i.e., the environment in which the probe moves) is affected by the intrinsic geometry as well as the topology of the probe.



\ack{We acknowledge the i-Link cooperation programme of CSIC (project i-Link0484) and thank the organizers of the Seventh Aegean Summer School at Parikia, Greece, where this work was born. 
 The work of GC is under a Ram\'on y Cajal contract.}

\

\end{document}